\documentclass{article}
\usepackage{spconf,amsmath,graphicx}
\usepackage{multirow}
\usepackage{bm}
\usepackage{amsfonts}
\usepackage[inkscapearea=page]{svg}
\usepackage{subfig}
\usepackage{tablefootnote}
\usepackage{url}
\usepackage[hidelinks]{hyperref}
\usepackage{amsmath}
\usepackage{textpos}

\newcommand{\R}{\mathbb{R}}
\newcommand{\tabindent}{\hspace{3mm}}

\newcommand{\din}{d_{\text{inter}}}
\newcommand{\Ymerge}{\bm{Y_{\text{Merge}}}}

\title{E-Branchformer: Branchformer with Enhanced merging \\ for speech recognition}
\name{Kwangyoun Kim$^1$, Felix Wu$^1$, Yifan Peng$^2$\sthanks{Work done during an internship at ASAPP.}, Jing Pan$^1$, Prashant Sridhar$^1$, Kyu J. Han$^1$, Shinji Watanabe$^2$}
\address{$^1$ASAPP Inc., Mountain View, CA, USA\\
$^2$Carnegie Mellon University, Pittsburgh, PA, USA\\}
\begin{document}
\ninept
\maketitle

\begin{abstract}
Conformer, combining convolution and self-attention sequentially to capture both local and global information, has shown remarkable performance and is currently regarded as the state-of-the-art for automatic speech recognition (ASR).
Several other studies have explored integrating convolution and self-attention but they have not managed to match Conformer's performance.
The recently introduced Branchformer achieves comparable performance to Conformer by using dedicated branches of convolution and self-attention and merging local and global context from each branch.
In this paper, we propose E-Branchformer, which enhances Branchformer by applying an effective merging method and stacking additional point-wise modules.
E-Branchformer sets new state-of-the-art word error rates (WERs) 1.81\% and 3.65\% on LibriSpeech test-clean and test-other sets without using any external training data.

\end{abstract}
\begin{keywords}
Automatic speech recognition, Conformer, Branchformer, Librispeech
\end{keywords}

\begin{textblock*}{\textwidth}(0cm, 12cm)
\tiny\noindent Copyright 2023 IEEE. Published in the 2022 IEEE Spoken Language Technology Workshop (SLT) (SLT 2022), scheduled for 19-22 January 2023 in Doha, Qatar. Personal use of this material is permitted. However, permission to reprint/republish this material for advertising or promotional purposes or for creating new collective works for resale or redistribution to servers or lists, or to reuse any copyrighted component of this work in other works, must be obtained from the IEEE. Contact: Manager, Copyrights and Permissions / IEEE Service Center / 445 Hoes Lane / P.O. Box 1331 / Piscataway, NJ 08855-1331, USA. Telephone: + Intl. 908-562-3966.
\end{textblock*}

\section{Introduction}
Automatic speech recognition (ASR) is a speech-to-text task, critical to enable spoken language understanding~\cite{feng2021asr,bastianelli2020slurp,shon2021slue}.
Recently, there has been a growing interest in end-to-end (E2E) ASR models for their simplicity, efficiency, and competitive performance. Thanks to many improvements in E2E ASR modeling such as new model architectures~\cite{chan2015listen,Graves2012SequenceTW,zhang2020transformer}, training objectives~\cite{chan2015listen,Graves2012SequenceTW,graves2006connectionist,kim2017joint}, and data augmentation methods~\cite{ko2015audio,park2019specaugment,wang2019semantic}, E2E ASR continues to take over conventional hybrid ASR in various voice recognition applications.

An acoustic encoder that extracts features from audio inputs plays a vital role in all E2E ASR models regardless of which training objective is applied to optimization. Recurrent neural networks~\cite{chan2015listen,rao2017exploring,zeyer2018improved,kim2019attention} used to be the de facto model choice for an encoder. Later, many convolution-based~\cite{li2019jasper,fujita2020attention,han2020contextnet} and Transformer-based~\cite{zhang2020transformer,dong2018speech, karita2019comparative,moritz2020streaming} models have also been introduced. The strength of multi-head self-attention in capturing global context has allowed the Transformer to show competitive performance. Transformers and their variants have been further studied to reduce computation cost~\cite{choromanski2020rethinking, wang2020linformer, zhai2021attention, tay2022efficient, wu2021fastformer} or to train deep models stably~\cite{wang2022deepnet}. 
In parallel, several studies have investigated how to merge the strengths of multi-head self-attention with those of other neural units. The combination of a recurrent unit with self-attention, which does not require explicit positional embeddings, shows stable performance even in long-form speech recognition~\cite{pan2021sru++}. Besides, Conformer~\cite{gulati2020conformer} which combines convolution and self-attention sequentially, and Branchformer~\cite{peng2022branchformer} which does the combination in parallel branches, both exhibit superior performance to the Transformer by processing local and global context together.

In this work, we extend the study of combining local and global information and propose \textit{E-Branchformer}, a descendent of Branchformer which enhances the merging mechanism. 
Our enhanced merging block utilizes one additional lightweight operation, a depth-wise convolution, which reinforces local aggregation.
Our experiments show that E-Branchformer surpasses Conformer and Branchformer on both test-clean and test-other in LibriSpeech.
Our specific contribution includes:
\begin{enumerate}
    \item The first Conformer baseline implementation using the attention-based encoder-decoder model that matches the accuracy of Google's Conformer~\cite{gulati2020conformer}.
    \item An improved Branchformer baseline that has a 0.2\% and 0.5\% improvement in absolute WER.
    \item An extensive study on various ways to merge local and global branches.
    \item The new E-Branchformer encoder architecture which sets a new state-of-the-art performance (1.81\% and 3.65\% WER on test-clean and test-other) on LibriSpeech under the constraint of not using any external data.
    \item Our code will be released at {https://anonymized\_url} for reproducibility.
\end{enumerate}

\section{Related Work}
\subsection{End-to-end Automatic Speech Recognition Models}
E2E ASR models can be roughly categorized into three main types based on their training objectives and decoding algorithms: connectionist temporal classification models~\cite{graves2006connectionist}, transducer models~\cite{Graves2012SequenceTW}, and the attention-based encoder-decoder (AED) models, a.k.a. Listen, Attend, and Spell (LAS) models~\cite{chan2015listen}. 

Regardless of the training objective and high-level framework, almost all E2E ASR models share a similar backbone; an acoustic encoder which encodes audio features into a sequence of hidden feature vectors. Many studies have explored different modeling choices for the acoustic encoder. Recurrent neural networks~\cite{chan2015listen,Graves2012SequenceTW,rao2017exploring,zeyer2018improved,hannun2014deep} encode audio signals sequentially; convolution alternatives~\cite{li2019jasper,fujita2020attention,han2020contextnet} process local information in parallel and aggregate them as the network goes deeper; self-attention~\cite{zhang2020transformer,dong2018speech, karita2019comparative,moritz2020streaming} allows long-range interactions and achieves superior performance.
In this work, we focus only on applying the E-Branchformer encoder under the AED framework, but it can also be applied to CTC or transducer models like the Conformer.

\subsection{Combining Self-attention with Convolution}
Taking the advantages of self-attention and convolution to capture both long-range and local patterns has been studied in various prior works. They can be categorized into two regimes: applying them sequentially or in parallel (i.e., in a multi-branch manner).

\subsubsection{Sequentially}
To the best of our knowledge, QANet~\cite{yu2018qanet} for question answering is the first model that combines convolution and self-attention in a sequential order. QANet adds two additional convolution blocks (with residual connections) before the self-attention block in each Transformer layer.
Evolved Transformer~\cite{so2019evolved} also combines self-attention and convolution in a sequential manner.
In computer vision, non-local neural networks~\cite{wang2018non} also show that adding self-attention layer after convolution layers enables the model to capture more global information and improves the performance on various vision tasks.
Recently, a series of vision Transformer variants that apply convolution and self-attention sequentially are also proposed including CvT~\cite{wu2021cvt}, CoAtNet~\cite{dai2021coatnet}, ViTAEv2~\cite{zhang2022vitaev2},  MaxVit~\cite{tu2022maxvit}.
In speech, Gulati et al.~\cite{gulati2020conformer} introduce Conformer models for ASR and show that adding a convolution block after the self-attention block achieves the best performance compared to applying it before or in parallel with the self-attention.

\subsubsection{In parallel}
Wu et al.~\cite{wu2020lite} propose Lite Transformer using Long-Short Range Attention (LSRA) which applies multi-head attention and dynamic convolution~\cite{wu2019pay} in parallel and concatenates their outputs. Lite Transformer is more efficient and accurate than Transformer baselines on various machine translation and summarization tasks.
Jiang et al.~\cite{jiang2020convbert} extend this to large scale language model pretraining and introduce ConvBERT which combines multi-head attention and their newly proposed span-based dynamic convolution with shared queries and values in two branches. Experiments show that ConvBERT outperforms attention-only baselines (ELECTRA~\cite{clark2020electra} and BERT~\cite{devlin2018bert}) on a variety of natural language processing tasks.
In computer vision, Pan et al.~\cite{pan2022integration} share a similar concept except that they decompose a convolution into a point-wise projection and a shift operation.
In speech, Branchformer combines self-attention and the convolutional spatial gating unit (CSGU)~\cite{sakuma2021mlp} achieving performance comparable with the Conformer.
We will introduce more details in Section~\ref{sec:branchformer}.

\subsubsection{Hybrid - both Sequentially and in parallel}
Recently, Inception Transformer~\cite{si2022inception} which has three branches (average pooling, convolution, and self-attention) fused with a depth-wise convolution achieves impressive performance on several vision tasks. Our E-Branchformer shares a similar spirit of combing local and global information both sequentially and in parallel.

\section{Preliminary: BranchFormer} \label{sec:branchformer}
\begin{figure}[h]
\centering
\def\svgwidth{0.92\linewidth}
\graphicspath{{figs/}}
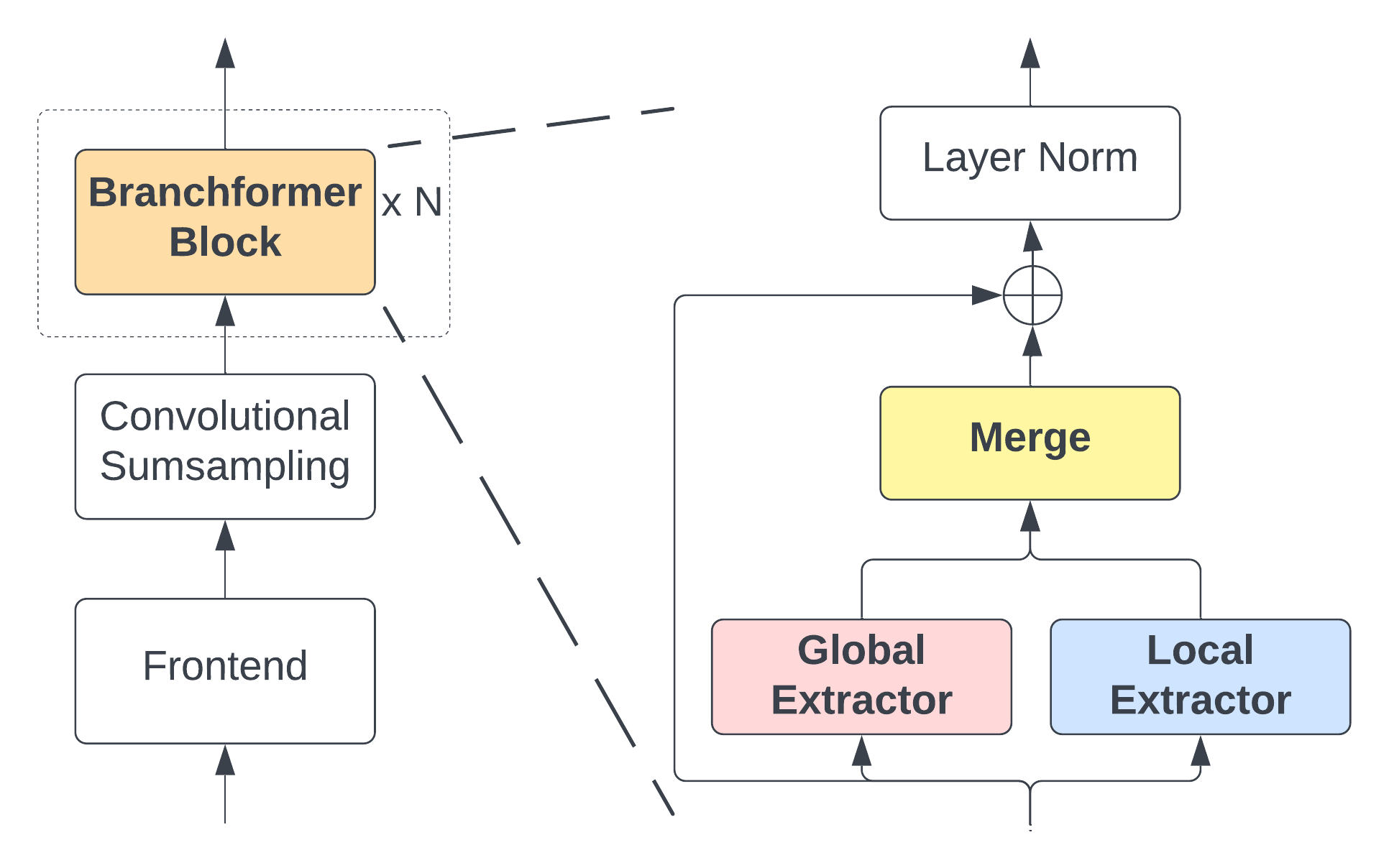
\caption[A figure of the Branchformer-based Encoder and a single Branchformer Block. There are two parallel branches to extract global and local context, and the merge module combines outputs of branches.]{A figure of the Branchformer-based Encoder and a single Branchformer Block. There are two parallel branches\footnotemark to extract global and local context, and the merge module combines outputs of branches.}
\label{fig:branchformer}
\end{figure}
\footnotetext{Unlike the Lite-Transformer and the in-parallel method in the Conformer, the Branchformer doesn't split the inputs for each branch along the channel dimension.}
Figure~\ref{fig:branchformer} shows the high-level architecture of the Branchformer encoder, which uses a frontend and a convolutional subsampling layer to extract low-level speech features and then applies several Branchformer blocks.
There are three components in each Branchformer block --- a global extractor branch, a local extractor branch, and a merge module.

The \textbf{global extractor branch} is a conventional self-attention block in Transformer.
It uses the pre-norm~\cite{he2016identity} setup where a layer norm (LN)~\cite{ba2016layer}, a multi-head self-attention (MHSA), and a dropout~\cite{srivastava2014dropout} are applied sequentially as follows:

\begin{equation}
    \begin{aligned}
        \bm{Y_G} = \mathrm{Dropout}(\mathrm{MHSA}(\mathrm{LN}(\bm{X}))),
    \end{aligned}
\end{equation}
where $\bm{X}, \bm{Y_G} \in \R^{T \times d}$ denote the input and the global-extractor branch output with a length of $T$ and a hidden dimension of $d$. 
Like the Conformer, Branchformer uses relative positional embeddings, which generally shows better performance than absolute positional embeddings in ASR and NLU tasks~\cite{gulati2020conformer,shaw2018self}. In the paper, the authors also explore more efficient attention variants to reduce computational cost, which leads to some degradation in accuracy.

\begin{figure}[h]
\centering
% \includesvg[width=0.6\linewidth]{figs/cgmlp.svg}
\def\svgwidth{0.72\linewidth}
\graphicspath{{figs/}}
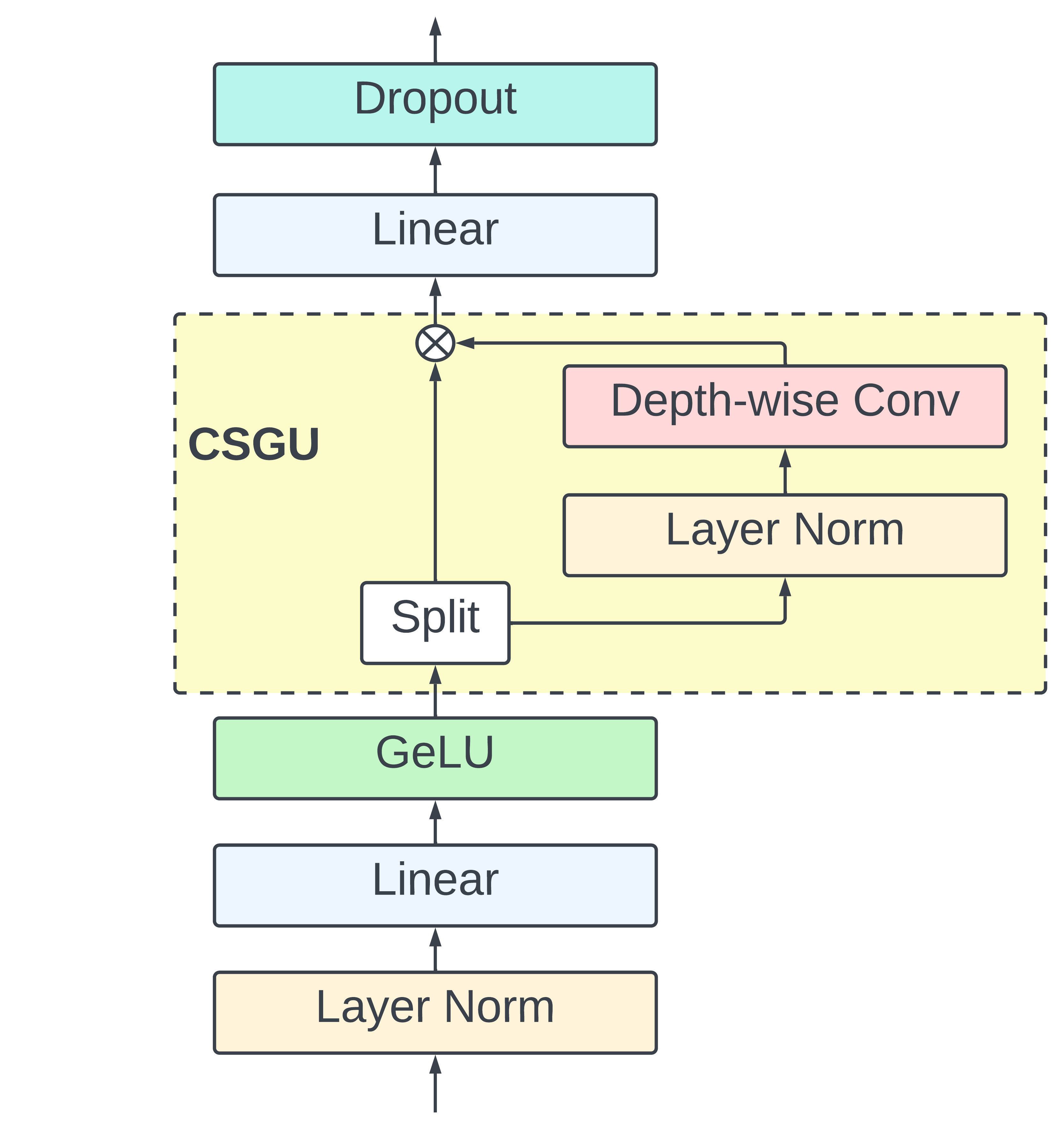
\caption{A figure of the Local extractor branch in Branchformer. This branch uses the Multi-Layer Perceptron (MLP) with convolutional gating (cgMLP)~\cite{sakuma2021mlp}.}
\label{fig:cgmlp}
\end{figure}
Figure~\ref{fig:cgmlp} illustrates the \textbf{local extractor branch} in Branchformer.
After LayerNorm, the input features are first projected to a higher dimension $d_{\mathrm{inter}}$ (usually $d_{\mathrm{inter}} = 6d$) followed by a GELU activation~\cite{hendrycks2016gaussian}.
Then, a Convolutional Spatial Gating Unit (CSGU)~\cite{sakuma2021mlp} is applied to encode local information. The CSGU block first splits the features into two parts (each with $\din / 2$ dimensions), then applies LayerNorm and depth-wise convolution on one of them, and finally multiplies two branches element-wisely.
The output of CSGU is projected back to $d$ dimensions followed by dropout.
Precisely, the local extractor branch processes input $\bm{X}$ and generates $\bm{Y_L}$ as follows:
\begin{align}
    &\bm{Z} = \mathrm{GELU}(\mathrm{LN}(\bm{X})\bm{U}), \\
    &[\bm{A}\;\bm{B}] = \bm{Z},  \\
    &\bm{\Tilde{Z}} = \mathrm{CSGU}(\bm{Z}) = \bm{A} \odot \mathrm{DwConv}(\mathrm{LN}(\bm{B})), \\
    &\bm{Y_L} = \mathrm{Dropout}(\Tilde{\bm{Z}}\bm{V}),
\end{align}
where $\bm{Z} \in \R^{T \times \din}, \bm{A}, \bm{B}, \bm{\Tilde{Z}} \in \R^{T \times \din / 2}$ are 
intermediate hidden features, $\odot$ is element-wise product, and $\bm{U} \in \R^{d \times \din}, \bm{V} \in \R^{\din / 2 \times d}$ denote the trainable weights of two linear projections. The local extractor branch outputs $\bm{Y_L }\in \R^{T \times d}$.   

In the \textbf{merge module}, Branchformer simply concatenates two outputs $\bm{Y_G}$ and ${\bm{Y_L}}$ and then applies a linear projection to reduce the dimensionality back to $d$:
\begin{equation}
    \begin{aligned}
    \Ymerge = \mathrm{Concat}(\bm{Y_G}, \bm{Y_L})\bm{W},
    \end{aligned}
\end{equation}
where $\bm{W} \in \R^{2d \times d}$ is the trainable weights of the linear projection. 

Alternatively, a weighted average method can be applied, which uses a scale for each branch and sum them up.
\begin{equation}
    \begin{aligned}
    \Ymerge = w_g\bm{Y_G} + w_l\bm{Y_L},
    \end{aligned}
\end{equation}
where $w_g, w_l \in R$ are the weights for the global and the local branch, respectively. 
The attention-based pooling can be applied to compute the weights, $w_g, w_l$. It is also possible to arbitrarily adjust the weight of each branch. A branch's weight can also be set to 0 to exclude the corresponding branch to reduce the amount of computation or for other purposes, which is called a branch drop. In addition, the weighted average method can be used to share a more reliable and interpretable analysis, such as the importance of each branch at a specific layer, by comparing and plotting the weight values. But, in terms of the accuracy, experiments show that the concatenation method is simpler but more accurate.

\section{E-BranchFormer}

\begin{figure*}[!t]
\centering
\subfloat[]{
% \includesvg[width=0.18\linewidth]{figs/wFFN_new.svg}
\def\svgwidth{0.18\linewidth}
\graphicspath{{figs/}}
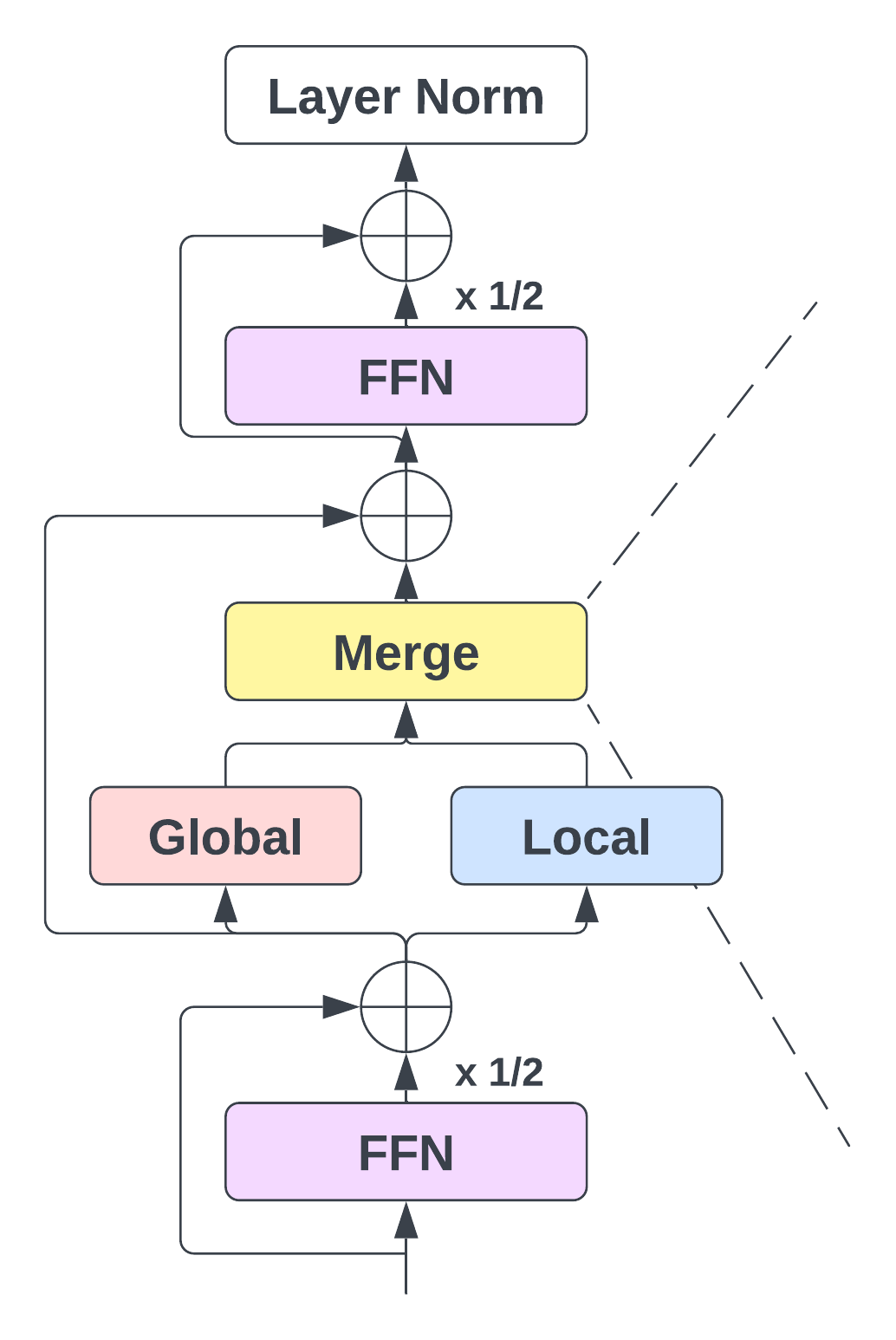
\label{fig:e_branchformer}
}
\subfloat[]{
% \includesvg[width=0.12\linewidth]{figs/merge_org.svg}
\def\svgwidth{0.12\linewidth}
\graphicspath{{figs/}}
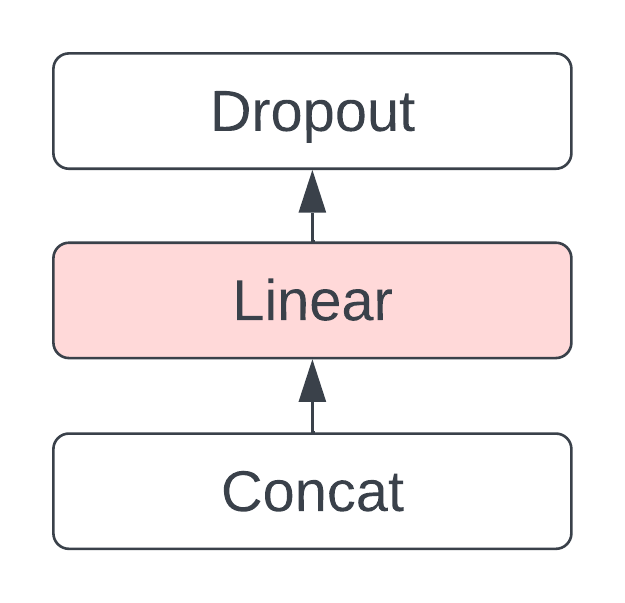
\label{fig:merge_org}
}
\subfloat[]{
% \includesvg[width=0.12\linewidth]{figs/merge_depth.svg}
\def\svgwidth{0.12\linewidth}
\graphicspath{{figs/}}
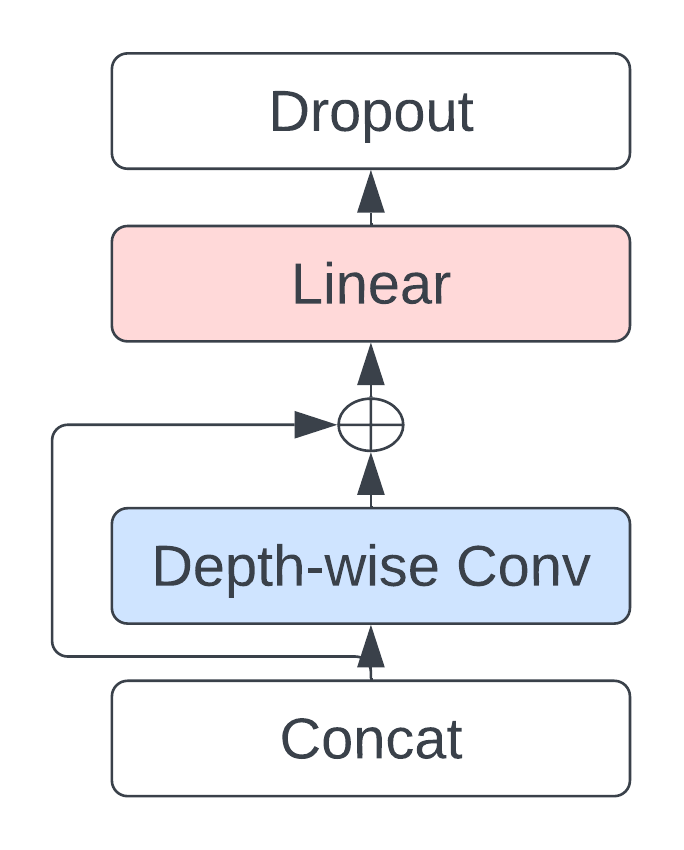
\label{fig:merge_depth}
}
\subfloat[]{
% \includesvg[width=0.12\linewidth]{figs/multi-kernel.svg}
\def\svgwidth{0.12\linewidth}
\graphicspath{{figs/}}
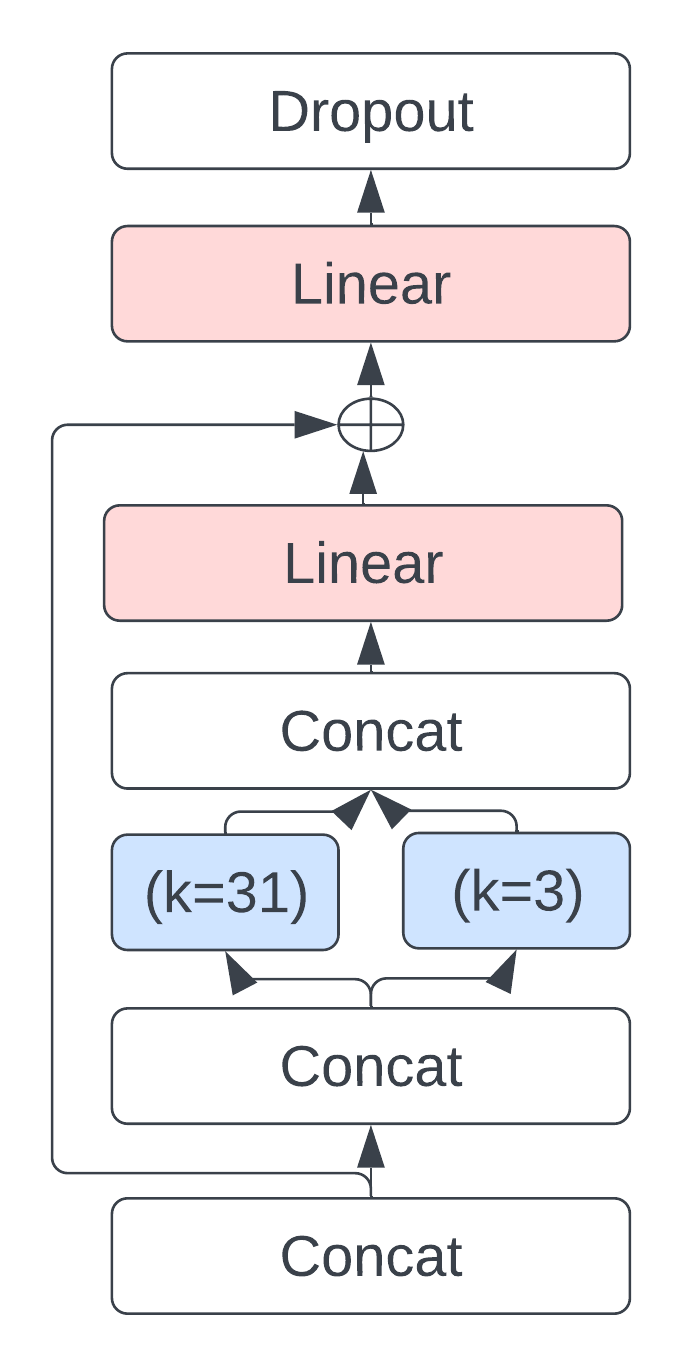
\label{fig:merge_multi}
}
\subfloat[]{
% \includesvg[width=0.12\linewidth]{figs/squeeze.svg}
\def\svgwidth{0.12\linewidth}
\graphicspath{{figs/}}
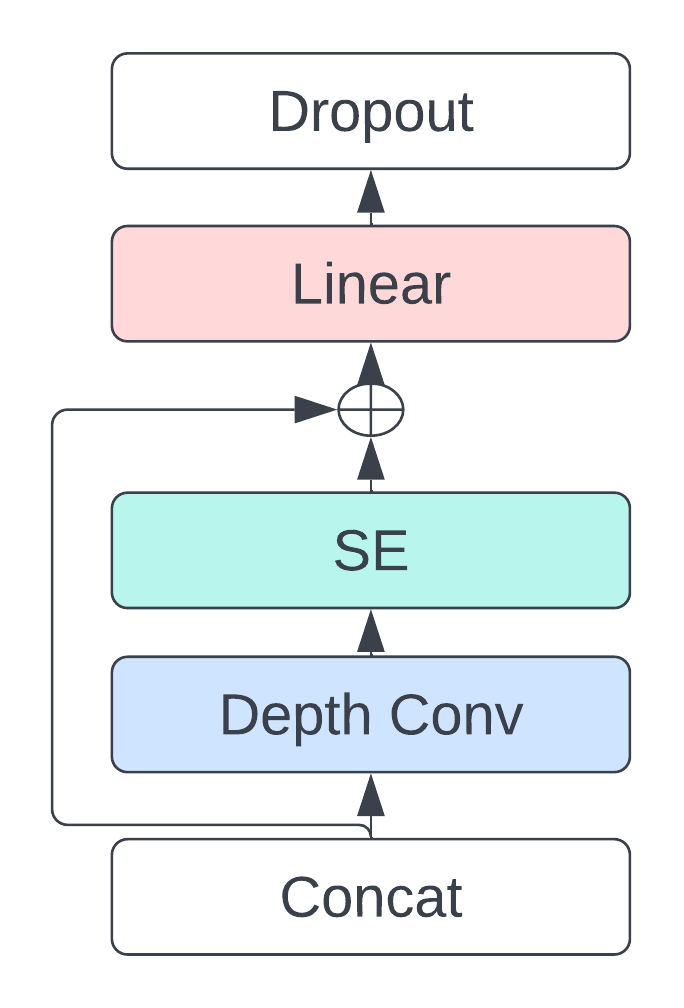
\label{fig:merge_squeeze}
}
\subfloat[]{
% \includesvg[width=0.12\linewidth]{figs/conv-module.svg}
\def\svgwidth{0.12\linewidth}
\graphicspath{{figs/}}
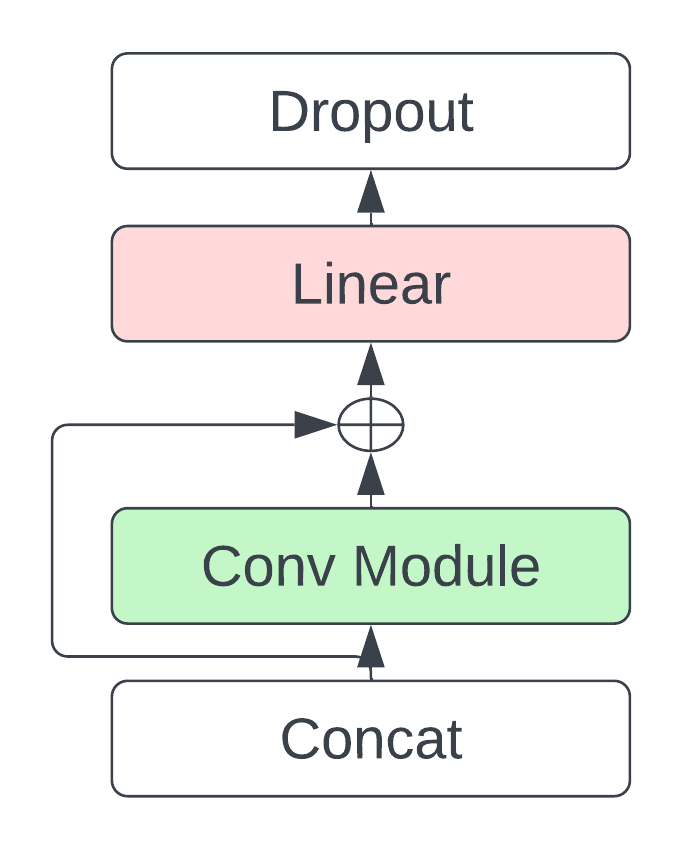
\label{fig:merge_conv_module}
}
\caption{A figure of (a) the proposed E-Branchformer block and different methods for the merge module: (b) is used in Branchformer, \textbf{(c)} is applying a depth-wise convolution, and (d) uses multiple convolutions with different kernel size, e.g., 31 and 3. (e) employs a squeeze-and-excitation block, and (f) replaces a depth-wise convolution with the convolution module in Conformer~\cite{gulati2020conformer}.}
\label{fig:merge_methods}
\end{figure*}

\subsection{Enhanced Merge Module}

We argue that it is suboptimal to combine the outputs of the two branches in the Branchformer point-wise and linearly. In this work, we study several potential modifications of the merging module that take temporal information into account. With such merge modules, self-attention and convolution would be combined sequentially and in parallel. Figure~\ref{fig:merge_methods} depicts the original merge module and five enhanced versions.

\subsubsection{Depth-wise Convolution}
We introduce depth-wise convolution to the merge module allowing it to take adjacent features into account when combing information from two branches. Notably, a depth-wise convolution requires little computation and has a negligible effect on the speed of the model.
% As described in Fig~\ref{fig:merge_org}, 
The original merge module in Branchformer concatenates outputs from the global and the local extractor and performs a linear projection along the feature dimension. In other words, the globalized and the localized context are computed in parallel and fused only across channels (as shown in Figure~\ref{fig:merge_org}). 
Presumably, using nearby information can improve the merge process.
Similar to Inception-Transformer~\cite{si2022inception}, we employ a depth-wise convolution to add the spatial information exchanging (as described in Figure~\ref{fig:merge_depth}). Formally, the outputs from the global $\bm{Y_G}$ and the local $\bm{Y_L}$ branch are merged:
\begin{align}
    &\bm{Y_{C}} = \mathrm{Concat}(\bm{Y_G}, \bm{Y_L}), \\
    & \bm{Y_D} = \mathrm{DwConv}(\bm{Y_C}), \label{eq:dwconv}\\
    &\Ymerge = (\bm{Y_C} + \bm{Y_D})\bm{W},
\end{align}
where $\mathrm{DwConv}$ stands for the depth-wise convolution and $\bm{W} \in \R^{2d \times d}$ is the trainable weights of the linear projection. 

Originally, the Inception-Transformer uses pooling and upsampling to reduce the amount of computation for the multi-head self-attention, which has quadratic time complexity with respect to the input sequence length. However, those pooling and upsampling operations make spatial information excessively smooth. Hence, when they combine the outputs of branches they utilize the depth-wise convolution to refine them. 
Despite not having such a smoothing issue, we discover that adding depth-wise convolution to spatially refine the extracted information from two branches enhances the merge process.
Beside enlarging the kernel size, we also explore ways to obtain information from more diverse perspectives through an additional convolution branch with different kernel size in parallel~\cite{ding2022scaling,liu2022more} (as shown in Figure~\ref{fig:merge_multi}).

\subsubsection{Squeeze-and-Excitation}
We examine a Squeeze-and-Excitation (SE) block~\cite{han2020contextnet,hu2018squeeze} to utilize global information during the merge process. 
An SE block takes a global average pooling over the temporal dimension, and feeds it to a tiny two-layer Feed-Forward Network (FFN) to produce a channel-wise gate.
Precisely,
\begin{align}
    & \bm{\bar{y}_{D}} = \frac{1}{T} \sum_{t=1}^T {\bm{Y_D}}_t,\\
    & \bm{g} = \sigma ( \mathrm{MLP}(\bm{\bar{y}_{D}}) ) = \sigma (\mathrm{Swish}(\bm{\bar{y}_{D}} \bm{W_1})) \bm{W_2}), \\
    &{\bm{Y'_D}}_i = \bm{g}_{i} \odot {\bm{Y_D}}_i \quad \forall i \in \{1, \dots, d\}, \\
    &\Ymerge = (\bm{Y_C} + \bm{Y'_D})\bm{W},
\end{align}
where $\bm{Y_D}$ is of the output of depth-wise convolution in Eq.~\ref{eq:dwconv}, $\mathrm{Swish}$ and $\sigma$ are Swish~\cite{ramachandran2017searching} and sigmoid non-linearity, respectively, $\odot$ denotes channel-wise multiplication, and $\bm{W_1} \in \R^{d \times d/8}$ and $\bm{W_2} \in \R^{d/8 \times d}$ are the trainable weights of the two-layer MLP. We omit the bias terms in all linearly transformation for simplicity.
The outputs of depth-wise convolution $\bm{Y_D}$ are now gated by the global information to produce the merge output $\Ymerge$.

\subsection{Revisiting the Point-wise Feed-Forward Network}
Since Branchformer has two linear projection layers in the cgMLP-base local extractor branch, for computational efficiency, it does not have FFN blocks like Transformer and Conformer. 
However, the role of the linear projections in the cgMLP and the FFNs could be different --- the FFNs in Transformer are used to refine pointwise information after temporal features are aggregated. 

Transformer and its variants commonly stack multi-head self-attention and FFN modules in an interleaving pattern. According to a study ~\cite{press2020improving}, models trained by stacking in random order sometimes outperform the interleaving pattern. They also conduct experiments to intentionally stack a specific module on the bottom or top in order to achieve better performance. In other words, the multi-head attention and the FFNs may not necessarily be stacked in interleaving patterns. However, they also say that using a balanced number of multi-head self-attention and FFN modules is not necessary, but it is generally desirable. In our case, it is possible that stacking E-Branchformer with FFN modules may perform better than stacking only E-Branchformer. In this direction, when building the Encoder, we revisit the FFNs or the macaron-style FFNs with $4d$ intermediate dimension and stack them together with E-Branchformer in an interleaving pattern to increase the expected model capacity. 
\section{Experiments}

\subsection{Experimental Setups}
We conduct experiments mainly on the LibriSpeech dataset~\cite{panayotov2015librispeech}, which contains approximately 1000 hours of transcribed speech and an additional text-only corpus mainly used for the external language model. 

ESPnet~\cite{watanabe2018espnet} is used for all experiments. ESPnet, an open source toolkit, has been widely used in many publications, and most paper experimental results are released as recipes or downloadable models. We utilize this to perform more accurate comparisons with existing methods.

We employ the attention-based encoder-decoder model (AED) as our base structure for all experiments. We use 80 dimensional log Mel features as input extracted with a 32ms window size and a 10ms stride, and 5K BPE sub-word units as output tokens.
At the bottom of the encoder is a convolutional subsampling module consisting of two 2D convolution layers, a ReLU and a linear layer. Each convolution layer uses a (3x3) kernel and a (2x2) stride effectively subsampling by a factor of 4, and the channel size of each convolution layer is the same as the hidden dimension of the encoder.
Our proposed E-Branchformer blocks are stacked on top of the subsampling module. We consider two main model sizes, Base and Large, and parameters are adjusted according to the experiment. The default Base encoder consists of 16 layers. For each layer, we use a feature hidden dimension $d$ of 256. The default Large encoder has 17 layers with a feature dimension of 512. The number of heads in the self-attention is $d/64$ and the intermediate hidden dimension of the cgMLP and the FFN are $6d$ and $4d$, respectively. When applying the macaron-style FFN, we reduce the number of layers or narrow down the intermediate hidden dimension of the FFN as $2d$ to fit the model size. For the encoder, we use a dropout rate and a layer dropout rate of 0.1. We use the swish activation in FFNs and convolution modules, and the kernel size of the convolution modules is 31. 
For all experiments, we employ a 6-layer Transformer decoder with the same feature hidden dimension $d$ as in the encoder. The dropout rate is adjusted to 0.2. 

\subsection{Training}
We use 8 NVIDIA A100 GPUs for model training, and each training is finished within 3 days for large models.
During training, we apply SpecAug~\cite{park2019specaugment} using 2 frequency-masks with parameter $F$ of 27, and 10 time-masks with parameters $p$ of 0.05. In addition, we augment our dataset by Speed Perturbation with scaling factors of 0.9, 1.0, and 1.1.
We employ a joint CTC-attention loss function with a CTC loss weight of 0.3. A label smoothing weight of 0.1 and the Adam optimizer with a weight decay of $10^{-6}$ are also utilized.
We apply the learning rate scheduling scheme named \textit{warmuplr} already implemented in ESPnet, which has a warm-up and a decay stage. We use 40K steps for the warm-up stage, and the model is trained for 80 epochs which are about 220K steps.
We select the final model by averaging 10 checkpoints with the highest validation accuracy. Since we train and test models using online-published recipes, most base experimental results are easily reproducible. We also use the automatic mixed-precision training implemented in PyTorch. 

\subsection{Inference}
We employ joint CTC-attention decoding with tuned weight.
In addition, we use an external language model (LM) for shallow fusion. It is a Transformer-based model with 16 layers, 128 embedding dimensions, 512 attention dimensions and 8 attention heads - in total 53.71M parameters. The recipe along with the pre-trained language model can be found in the ESPnet toolkit\footnote{\href{https://github.com/espnet/espnet/tree/master/egs2/librispeech/asr1}{https://github.com/espnet/espnet/tree/master/egs2/librispeech/asr1}}.

To further improve the language model shallow fusion and boost accuracy, we apply the \textbf{Internal Language Model Estimation (IMLE)} method proposed by Meng et al.~\cite{meng2021internal}. It is known that E2E ASR models implicitly learn an internal language model (ILM) over the training text distribution. Estimating the internal language model distribution allows us to explicitly interpolate internal LM and external LM with tunable hyperparameters, resulting in more effective decoding. At each decode step, the ILME follows the same AED decode step except that the source attention is disabled by an attention mask --- effectively the decoder estimates the probability over tokens solely based on the text. For each hypothesis, the score is given by
\begin{equation*}
\begin{aligned}
\log(P(Y)) = & \log(P(Y|X; \theta^{\mathrm{AED}})) \\ 
              & - \lambda_{\mathrm{ilm}}\log(P(Y; \theta^{\mathrm{AED}})) + \lambda_{\mathrm{elm}}\log(P(Y; \theta^{\mathrm{LM}})),
\end{aligned}
\end{equation*}
where $\lambda_{\mathrm{ilm}}$ and $\lambda_{\mathrm{elm}}$ are the interpolation weights for the internal and external langauge model, respectively. $P(Y|X; \theta^{\mathrm{AED}})$ is the probability of the hypothesis $Y$ yielded by the ASR model given the input acoustic feature $X$. $P(Y; \theta^{\mathrm{AED}})$ and $P(Y; \theta^{\mathrm{LM}})$ represent the internal language model estimation and the external language model probability over the hypothesis. In our experiment, we set $\lambda_{\mathrm{ilm}}$ and $\lambda_{\mathrm{elm}}$ to 0.2 and 0.6, respectively.

\section{Result}

\subsection{Main result}
% \begin{table}[th]
% \caption{Comparing word error rates (WER) with other published models. The external LM used in \cite{guo2021recent, pan2021sru++, peng2022branchformer} is the same as what we used in all our experiments.}
% \label{table:main_result}
% \begin{center}
% \resizebox{\linewidth}{!}
% {
% \begin{tabular}{lccccc}
% \hline
% \multirow{3}{*}{Model} & \multirow{2}{*}{Params} & \multicolumn{2}{c}{Without LM} & \multicolumn{2}{c}{With LM} \\
%  & & test & test & test & test \\
%  & (M) & clean & other & clean & other \\
% \hline\hline
% \textbf{Transducer} & & & & & \\
% Transformer\cite{zhang2020transformer} & 139 & 2.4 & 5.6 & 2.0 & 4.6 \\
% ContextNet\cite{han2020contextnet} & 112.7 & 2.1 & 4.6 & 1.9 & 4.1 \\
% Conformer\cite{gulati2020conformer} & 118.8 & 2.1 & 4.3 & 1.9 & 3.9 \\
% \hline
% \textbf{LAS} & & & & & \\
% Transformer\cite{synnaeve2019end} & 270 & 2.9 & 7.0 & 2.3 & 5.2 \\
% Conformer\cite{guo2021recent} & 116.15 & - & - & 2.1 & 4.9 \\
% SRU++\cite{pan2021sru++} & 114.61 & - & - & 2.0 & 4.7 \\
% Branchformer\cite{peng2022branchformer} & 116.2 & 2.4 & 5.5 & 2.1 & 4.5 \\
% \hline
% \textbf{Our baselines (LAS)} & & & & & \\
% Conformer & 147.77 & - & - & - & - \\
% Branchformer & 142.42 & 2.2 & 5.1 & 1.9 & 4.0 \\
% \hline
% \textbf{Our works (LAS)} & & & & & \\
% E-Branchformer(B) & - & - & - & - & - \\
% E-Branchformer(L) & 148.41 (115.98) & 2.1 & 4.8 & 1.8 & 3.8 \\
% \tabindent{+ ILME} & - & - & - & 1.7 & 3.7 \\
% \hline
% \end{tabular}
% }
% \end{center}
% \end{table}
\begin{table}[th]
\caption{Comparing word error rates (WER) with other published models. The external LM used in~\cite{pan2021sru++, peng2022branchformer,guo2021recent} is downloadable from ESPnet, and it is the same as what we used in all our experiments.}
\label{table:main_result}
\begin{center}
\resizebox{\linewidth}{!}
{
\begin{tabular}{lcccccc}
\hline
\multirow{3}{*}{Model} & \multirow{2}{*}{Params} & \multirow{2}{*}{Enc.} & \multicolumn{2}{c}{Without LM} & \multicolumn{2}{c}{With LM} \\
 & & & test & test & test & test \\
 & (M) & (M) & clean & other & clean & other \\
\hline\hline
\textbf{Transducer} & & & & & & \\
Transformer~\cite{zhang2020transformer} & 139 & - & 2.4 & 5.6 & 2.0 & 4.6 \\
ContextNet~\cite{han2020contextnet} & 112.7 & - & 2.1 & 4.6 & 1.9 & 4.1 \\
Conformer (M)~\cite{gulati2020conformer} & 30.7 & 27.3 & 2.3 & 5.0 & 2.0 & 4.3 \\
Conformer (L)~\cite{gulati2020conformer} & 118.8 & 114.9\tablefootnote{This is the estimated size based on the information in the paper~\cite{gulati2020conformer}} & 2.1 & 4.3 & 1.9 & 3.9 \\
\hline
\textbf{AED} & & & & & & \\
Transformer~\cite{synnaeve2019end} & 270 & - & 2.9 & 7.0 & 2.3 & 5.2 \\
Conformer~\cite{guo2021recent} & 116.2 & 83.2 & - & - & 2.1 & 4.9 \\
SRU++~\cite{pan2021sru++} & 114.6 & - & - & - & 2.0 & 4.7 \\
Branchformer~\cite{peng2022branchformer} & 116.2 & 83.3 & 2.4 & 5.5 & 2.1 & 4.5 \\
% \hline
% \textbf{Our baselines (AED)} & & & & & & \\
% Conformer & 147.8 & 114.9 & - & - & - & - \\
% Branchformer & 142.4 & 113.8 & 2.2 & 5.1 & 1.9 & 4.0 \\
% \hline
% \textbf{Our works (AED)} & & & & & & \\
% E-Branchformer(B) & - & - & - & - & - \\
% E-Branchformer(L) & 148.4 & 116.0 & 2.1 & 4.8 & 1.8 & 3.8 \\
% \tabindent{+ ILME} & - & - & - & - & 1.7 & 3.7 \\
% \hline
\hline
\textbf{Our baselines (AED)} & & & & & & \\
Conformer & 147.8 & 114.9 & 2.16 & 4.74 & 1.84 & 3.95 \\
Branchformer & 146.7 & 113.8 & 2.25 & 4.83 & 1.93 & 4.00 \\
\hline
\textbf{Our works (AED)} & & & & & & \\
E-Branchformer (B) & 41.12 & 27.8 & 2.49 & 5.61 & 1.97 & 4.26 \\
% E-Branchformer (L) & 148.4 & 115.5 & 2.11 & 4.76 & 1.78 & 3.82 \\
% \tabindent{+ ILME} & - & - & - & - & 1.74 & 3.74 \\
E-Branchformer (L) & 148.9 & 116.0 & 2.14 & 4.55 & 1.85 & 3.71 \\
\tabindent{+ ILME} & - & - & - & - & 1.81 & 3.65 \\
\hline
\end{tabular}
}
\end{center}
\end{table}
We compare the word error rates (WER) of our proposed E-Branchformer with those of previous studies. Models are evaluated on test-clean and test-other, with or without an external LM. 
First, we note that the Transducer-based models in Table~\ref{table:main_result}, especially Conformer-Transducer~\cite{gulati2020conformer}, known as state-of-the-art in terms of an accuracy, have not published a model or open sourced their training recipes and we have not been able to reproduce their results. Instead of reproducing them, we use only the WER and information presented in each paper. In order to compare relative performance with Conformer, we perform experiments on AED models and by changing the Encoder structure.

As shown in Table~\ref{table:main_result}, Branchformer~\cite{peng2022branchformer} performs 0.4\% better than Conformer~\cite{guo2021recent} on test-other with a LM. However, it is still worse than Conformer-Transducer by 0.2\% and 0.6\% on test-clean and test-other, respectively. As mentioned earlier, it is difficult to directly compare the two studies because the base code, the structure of the decoder, and the external LM are all different. However, from comparison of studies, we find that the encoder of Conformer-Transducer has more parameters than the encoder of Branchformer-AED or Conformer-AED.

In order to fairly compare AED models, we configure the encoder of AED-based models following the encoder of Conformer-Transducer. In the case of the Conformer-AED model, we use the configuration as described in ~\cite{gulati2020conformer}, and in the case of Branchformer, we stack more blocks in order to achieve a similar encoder size. The overall model size varies due to the decoder's structure, but encoder size is similar to that of the Conformer-Transducer. As a result, we obtain significantly improved performance compared to the existing AED-based models. 
By decoding Branchformer with shallow fusion, we achieve a WER of 1.93\% on test-clean and 4.0\% on test-other. Similarly, we obtain 1.84\% and 3.95\% by using Conformer-AED, which is very similar to the accuracy of the existing Conformer-Transducer. We use these as new baselines.

Then, we train the proposed E-Branchformer model and obtain performance better than that of existing studies and our baselines. In particular, with the E-Branchformer (L) model, which exploits depth-wise convolution with a kernel size of 31 in the merge module as illustrated in Figure ~\ref{fig:merge_depth} and the narrowed macaron-style FFN, with the external LM, we obtain WER of 1.85\% and 3.71\% on test-clean and test-other, respectively. This is better than the 1.9\% and 3.9\% of the existing state-of-the-art in WER. Compared to Branchformer~\cite{peng2022branchformer} which has much larger size, our E-Branchformer (B), which uses the depth-wise convolution in the merge module and a basic FFN, shows similar performance without the LM, and outperforms on both test-clean and test-other by 0.1\% and 0.2\%, respectively, with the LM.

Subsequently, we employ Internal Language Model Estimation (ILME) to further enhance the effect of the shallow fusion. This finally leads to the remarkable result of 1.81\% and 3.65\% on test-clean and test-other respectively, which is a new state-of-the-art when no external data is used\footnote{W2v-BERT~\cite{chung2021w2v} achieves 1.4\% and 2.5\% WER using additional 60K hours of unlabelled data from LibriLight~\cite{kahn2020libri}.} as far as we know. Some notable and valuable model results through combinations of different methods are analyzed in Section~\ref{sec:ablation}.

\subsection{Ablation Study} \label{sec:ablation}

\subsubsection{FFN module}
\begin{table}[th]
\caption{Comparison of word error rates (WER) and the computational complexity (MACs) depending on whether FFN, macaron-style FFN (FFN-mac), or macaron-style FFN with the narrow intermediate dimension (FFN-mac-n) is applied or not. We use the Large size model without any change in the original Branchformer block. No LM is used for evaluations.}
\label{table:ffn}
\begin{center}
\resizebox{\linewidth}{!}
{
\begin{tabular}{lccccccc}
\hline
\multirow{2}{*}{Model} & \# of & Enc. & MACs & dev & dev & test & test \\
 & layers & Params (M) & (G) & clean & other & clean & other \\
\hline\hline
% BranchFormer & 24 & 142.42 & 2.1 & 5.0 & 2.2 & 5.0 \\
BranchFormer & 25 & 113.8 & 43.7 & 2.01 & 4.99 & 2.25 & 4.83 \\
\tabindent{+ FFN} & 17 & 115.4 & \textbf{42.6} & 1.98 & 4.83 & 2.18 & 4.87 \\
\tabindent{+ FFN-mac} & 13 & 117.3 & \textbf{42.3} & 1.97 & 4.74 & 2.26 & 4.74 \\
\tabindent{+ FFN-mac-n} & 17 & 115.5 & \textbf{42.6} & 1.97 & 4.78 & 2.10 & 4.79 \\
\tabindent{+ FFN-mac} & 17 & 151.1 & 51.5 & 2.00 & 4.75 & 2.15 & 4.47 \\
\hline
\end{tabular}
}
\end{center}
\end{table}
Table~\ref{table:ffn} shows the performance changes through cooperating with FFN or macaron-style FFN. In this experiment, we use the same structure of the Decoder and the same dimension of the encoder in all cases. We also adjust the number of stacked blocks accordingly to compare under similar encoder sizes.
17-layers of Branchformer with FFN performs better than the stack of 25-layers of Branchformer-only. In addition, we obtain better performance when we stack 13-layers of Branchformer with macaron-style FFN. To construct more deeper encoder, we narrow down the intermediate dimension for FFNs from $4d$ to $2d$ and stack 17-layers Branchformer without increasing the model size. And this model shows the best performance on average among the models with similar model sizes. However, since there is not much difference between those models using FFNs, we employ these variants in the rest of experiments to find the optimal combination with other modules.

Although the model sizes are similar, the amount of computations may vary depending on the structure of the model and the number of layers. We measure the multiply-accumulate operations (MACs) to compare the computational complexity of each models. We utilize DeepSpeed~\footnote{\href{https://github.com/microsoft/DeepSpeed}{https://github.com/microsoft/DeepSpeed}} toolkit for profiling and use a randomly generated 10 seconds audio as a dummy input.

From these results, we can confirm that using FFN together is a reasonable way to expect better accuracy than deeply stacking only Branchformers. 

\subsubsection{Kernel sizes of the depth-wise convolution}
In this study, we compare the effects of different kernel sizes in the depth-wise convolution, stacked sequentially in the merge module. We already know that Conformer has achieved the best performance with the kernel size of 31 or 32 for the depth-wise convolution in the sequentially stacked convolution module~\cite{gulati2020conformer, guo2021recent}. However, we try to verify through experiments whether additionally stacked depth-wise convolution also needs the large-sized kernel, even though the local branch of Branchformer~\cite{peng2022branchformer} or our E-Branchformer is already having a depth-wise convolution with the kernel size of 31. As shown in Table~\ref{table:conv_kernel}, using 31 is the most accurate, especially on dev-other and test-other by a large margin. We use this kernel size in other experiments.
\begin{table}[th]
\caption{Comparison of word error rates (WER) according to different kernel sizes of the depth-wise convolution in the merge module. The Base size model with FFN is employed for this ablation study, and we decode each models without an external LM.}
\label{table:conv_kernel}
\begin{center}
\resizebox{\linewidth}{!}
{
\begin{tabular}{c|cccc}
\hline
Kernel size & dev-clean & dev-other & test-clean & test-other \\
\hline\hline
3 & 2.33 & 5.99 & 2.64 & 5.97 \\
15 & 2.40 & 5.89 & 2.69 & 5.87 \\
\textbf{31} & \textbf{2.25} & \textbf{5.68} & \textbf{2.49} & \textbf{5.61} \\
63 & 2.40 & 5.81 & 2.53 & 5.78 \\
\hline
\end{tabular}
}
\end{center}
\end{table}

\subsubsection{Merge module}
\begin{table}[th]
\caption{Effect of different convolution-based methods added in the merge module. (b)-(f) refers to each method in Figure~\ref{fig:merge_methods}. The Base size model with FFN modules is used without an external LM.}
\label{table:merge_method}
\begin{center}
\resizebox{\linewidth}{!}
{
\begin{tabular}{l|cccccc}
\hline
 \multirow{2}{*}{Method} & Enc. & MACs & dev & dev & test & test \\
 & (M) & (G) & clean & other & clean & other \\
\hline\hline
w/o Depth-wise Conv (b) & 27.5 & 10.8 & 2.34 & 5.99 & 2.57 & 5.79 \\
\hline
w Depth-wise Conv (c) & 27.8 & 10.8 & \textbf{2.25} & \textbf{5.68} & 2.49 & 5.61 \\
\tabindent{+ Multiple Kernel} (d) & 36.2 & 12.9 & 2.34 & 5.87 & 2.58 & 5.90 \\
\tabindent{+ SE block} (e) & 28.9 & 10.9 & 2.27 & 5.82 & \textbf{2.42} & \textbf{5.59} \\
Conv Module (internal) (f) & 40.4 & 14.0 & 2.66 & 6.40 & 2.84 & 6.35 \\
Conv Module (external) & 30.8 & 11.6 & 2.36 & 5.87 & 2.64 & 5.88 \\
\hline
\end{tabular}
}
\end{center}
\end{table}
As shown in Table~\ref{table:merge_method}, we perform comparative experiment on which convolution-based method is more effective to enhance the merge module. We compare various methods illustrated in Figure~\ref{fig:merge_methods}.
First, when we exploit the depth-wise convolution as shown in Figure~\ref{fig:merge_depth}, performance is evenly improved on all evaluation sets by 0.1\% to 0.3\%. Whereas adding another depth-wise convolution with kernel size of 3, as described in Figure~\ref{fig:merge_multi}, is harmful for all evaluation sets, and applying additional linear projection before each depth-wise convolution to reduce the dimension shows even worse result. 
A SE block in Figure~\ref{fig:merge_squeeze} helps to obtain a marginal improvement on test-sets, but not on dev-sets. We originally think that this gating-based mechanism reinforces the depth-wise convolution output by reflecting the global information, but we interpret this result that the effect may be relatively reduced due to the presence of channels from the global branch. Although the total computations is almost the same, it is difficult to say that the SE block plays a beneficial role in this experiment.
We also try to borrow the convolution module from Conformer, which cooperates with the self-attention module sequentially. But, when we internally replace the depth-wise convolution with it, as depicted in Figure~\ref{fig:merge_conv_module}, the performance is extremely degraded. The other way we use it is to stack outside of the merge module before FFN externally. This method performs better than when it is inside the merge module, but it still makes the model perform poorly. The reason, why it does not provide additional improvement, can be assumed to be the redundancy caused by the convolution module being more focused on extracting the local context like what our cgMLP is already doing. 

In conclusion, simply adding the depth-wise convolution is effective in terms of performance and also efficient in terms of the parameter size and the computational complexity.
\section{Conclusion}
We have proposed E-Branchformer, a new encoder architecture for automatic speech recognition. 
With an enhanced merging mechanism that allows hybrid application (both sequentially and in parallel) of self-attention and convolution, E-Branchformer outperforms both Conformer and Branchformer and sets a new-state-of-art on LibriSpeech test sets without using any external data.
We are interested in applying to other ASR models, such as Transducer, and exploring the potential of E-Branchformer on other speech tasks such as self-supervised learning, speech identification, speech enhancement, and spoken language understanding tasks in the future.

\bibliographystyle{IEEEbib}
\bibliography{refs}

\end{document}